# A Steel Surface Defect Detection Method Based on Lightweight Convolution Optimization

Cong Chen[1], Ming Chen[2], Hoileong Lee[3], Yan Li[4], and Jiyang YU[5]*
School of Marine Information Engineering, Hainan Tropical Ocean University, Sanya 572022, China[1]
School of Information and Intelligent Engineering, University of Sanya, Sanya 572022, China[2]
Faculty of Electronic Engineering & Technology, Universiti Malaysia Perlis, 02600 Arau, Perlis, Malaysia[3]
Thalesgroup, Ottawa, ON, K1K 4Z9, Canada[4]
Graduate School of Management of Technology, Pukyong National University, Busan 48547, Korea[5]

*Abstract*—Surface defect detection of steel, especially the recognition of multi-scale defects, has always been a major challenge in industrial manufacturing. Steel surfaces not only have defects of various sizes and shapes, which limit the accuracy of traditional image processing and detection methods in complex environments. However, traditional defect detection methods face issues of insufficient accuracy and high miss-detection rates when dealing with small target defects. To address this issue, this study proposes a detection framework based on deep learning, specifically YOLOv9s, combined with the C3Ghost module, SCConv module, and CARAFE upsampling operator, to improve detection accuracy and model performance. First, the SCConv module is used to reduce feature redundancy and optimize feature representation by reconstructing the spatial and channel dimensions. Second, the C3Ghost module is introduced to enhance the model's feature extraction ability by reducing redundant computations and parameter volume, thereby improving model efficiency. Finally, the CARAFE upsampling operator, which can more finely reorganize feature maps in a content-aware manner, optimizes the upsampling process and ensures detailed restoration of high-resolution defect regions. Experimental results demonstrate that the proposed model achieves higher accuracy and robustness in steel surface defect detection tasks compared to other methods, effectively addressing defect detection problems.

*Keywords—YOLOv9s; steel surface defect detection; C3Ghost module; SCConv module; CARAFE upsampling operator*

## I. INTRODUCTION

The surface quality of steel is one of the key indicators for measuring its performance and reliability, and it is widely applied in high-precision fields such as aerospace, automotive manufacturing, and building structures. Surface defects not only compromise the aesthetic quality of steel but also jeopardize its structural integrity, potentially causing safety hazards. Therefore, steel surface defect detection has become an important task in modern manufacturing. Traditional defect detection methods, such as manual visual inspection, ultrasonic testing, and optical scanning, although still of certain value in specific situations, face issues such as low efficiency, poor accuracy, and difficulty in identifying complex defects due to their reliance on manual operation or being limited to specific detection conditions [1]. In particular, in large-scale production environments, manual inspection cannot meet the requirements for high speed, high precision, and high reliability [2].

In recent years, with the rapid development of deep learning technology, particularly breakthroughs in the field of computer vision, automated defect detection technology has seen significant improvements. Deep learning algorithms, especially Convolutional Neural Networks (CNNs), have demonstrated powerful capabilities in image recognition, enabling efficient and accurate identification and classification of various defects on steel surfaces. Compared to traditional methods, deep learning-based detection methods not only improve detection efficiency but also overcome the limitations of manual inspection, adapting to the complex and variable production environment. In particular, the YOLO (You Only Look Once) [3-6] object detection algorithm, due to its efficient real-time performance and accuracy, has been widely adopted for steel defect detection.

Many researchers have made significant contributions to the field of defect detection. For instance, Gao et al. [7] integrated the attention mechanism and weighted bi-directional feature pyramid network (BiFPN) into the YOLOv5 architecture, achieving good results. However, there are still certain shortcomings when handling small-sized defects. Yu et al. [8] proposed the introduction of structural reparameterization, context transformation modules, and simplified generalized feature pyramid networks to improve the model's accuracy. Although they performed excellently in terms of mAP, complex background noise affected defect detection performance, indicating that there is still room for improvement in the model's robustness to interference. Yang et al. [9] proposed a detection method combined with a supervised spatial attention module (SSAM) to improve the detection accuracy of defects such as surface cracks and rolled-in scale on steel. However, this method increases the computational complexity of the model and raises hardware requirements. Subburaj et al. [10] proposed the DBCW-YOLO model, which integrates attention mechanisms and enhanced feature extraction techniques to improve defect detection accuracy. However, integrating multiple advanced technologies increases the model's complexity, affecting real-time performance. Wu et al. [11] proposed the Hyper-YOLO model, which improves detection performance by replacing the CSP module in YOLOv5 with the Ghost module, the PAFPN module with the Hyper FPN module, and introducing α-CIoU and α-DIoU loss functions. However, these improvements increase the model's computational cost. Xu et al. [12] improved the YOLOv5 algorithm by introducing a coordinate attention mechanism and multi-scale feature fusion, significantly improving the model's detection accuracy. However, these improvements increase the model's complexity, leading to longer training and inference times. Kong et al. [13] proposed an improved steel surface defect detection algorithm based on YOLOv8, enhancing small-target detection accuracy by introducing a non-attention mechanism and improving the SPPF module. However, the algorithm requires substantial computational resources, limiting its deployment in resource-constrained environments. Mi et al. [14] proposed a surface defect detection method for hot-rolled steel strips based on multi-scale feature perception and adaptive feature fusion, effectively improving the model's detection performance. However, it performs poorly when handling highly complex or noisy images and requires longer training times.

In summary, although significant progress has been made in existing steel surface defect detection algorithms, there are still numerous challenges and areas for optimization, especially regarding the detection accuracy of small-sized defects. To address this, this paper proposes a steel surface defect detection method based on deep learning with YOLOv9s, aiming to solve the problems encountered by current methods when handling small-sized defects through a series of optimization designs, thereby improving detection accuracy and efficiency. The core contributions of this study are mainly reflected in the following key aspects:

*1)* The SCConv module is adopted, and through the aforementioned reconstruction of spatial and channel dimensions, the model can reduce redundant information without losing important features, making the network more efficient and improving feature representation accuracy.

*2)* The C3Ghost module is introduced, which helps the model focus on the most distinguishable defect features by reducing redundant feature channels, while significantly decreasing the computational and memory load.

*3)* The CARAFE (Content-Aware Re-Assembly of Feature Maps) upsampling operator is incorporated, which, through content-based feature reorganization, dynamically adjusts the resolution of the feature map, thereby better preserving the spatial information of the image during the upsampling process.

## II. RELATED WORK

### A. Depthwise Separable Convolution

Depthwise separable convolution [15] is an efficient convolutional operation that significantly reduces computational cost and parameter count by decomposing the traditional convolution operation into two simpler operations. This approach enhances the efficiency of neural networks. The process consists of the following steps: In a standard convolution, the input feature map has a size of $H \times W \times M$, the convolution kernel has a size of $K \times K \times M$, and the number of convolution kernels is $N$, resulting in an output feature map of size $H \times W \times N$. Consequently, the parameter count of standard convolution is $P_1 = K \times K \times M \times N$, and the computational cost is $C_1 = H \times W \times K \times K \times M \times N$.

Depthwise separable convolution decomposes the standard convolution kernel into two parts: depthwise convolution and pointwise convolution. Depthwise convolution applies $K \times K \times 1$ convolution kernels to each input channel independently, with a total of $M$ such kernels. Pointwise convolution then uses $1 \times 1 \times M$ convolution kernels to combine all input channels, ultimately generating $N$ output channels. This approach significantly reduces computational cost and parameter count, enhancing network efficiency. Consequently, the total parameter count of depthwise separable convolution is $P_2 = K \times K \times M + M \times N$, with a computational cost of $C_2 = K \times K \times M \times H \times W + M \times N \times H \times W$. Specifically, the parameter count for depthwise convolution is $K \times K \times M$, and for pointwise convolution, it is $M \times N$. The computational cost for depthwise convolution is $K \times K \times M \times H \times W$, while for pointwise convolution, it is $M \times N \times H \times W$.

From the perspective of parameter count and computational cost, depthwise separable convolution significantly reduces both by decomposing the operations of standard convolution. The ratios of the parameter count and computational cost between depthwise separable convolution and standard convolution are $\frac{P_2}{P_1} = \frac{K \times K \times M + M \times N}{K \times K \times M \times N} = \frac{1}{N} + \frac{1}{K^2}$ and, $\frac{C_2}{C_1} = \frac{K \times K \times M \times H \times W + M \times N \times H \times W}{H \times W \times K \times K \times M \times N} = \frac{1}{N} + \frac{1}{K^2}$, respectively. It can be observed that depthwise separable convolution effectively lowers computational overhead, especially when processing high-dimensional inputs. Therefore, compared to standard convolution, depthwise separable convolution offers higher computational efficiency.

### B. YOLOv9 Algorithm

The YOLOv9 model features a unique design in its three key components: the backbone network, the neck network, and the head network. The backbone network adopts the General Efficient Layer Aggregation Network (GELAN), which ingeniously integrates CSPNet and ELAN. This fusion increases the network's width, facilitating smoother gradient flow throughout the network and enhancing its feature extraction capability. The neck network, serving as a crucial bridge between the backbone and head, employs the PGI-based multi-level auxiliary information module. This design enables each level of the feature pyramid to receive information on all object sizes, integrating gradient information from different prediction branches to facilitate parameter updates. As a result, it effectively achieves multi-scale feature fusion while incorporating both deep and shallow feature information without significantly increasing computational cost. The head network adopts a decoupled head design, which consists of two branches: a classification branch and a bounding box (box) branch. Additionally, it divides detection boxes into three scales—large, medium, and small—targeting objects of different sizes. The classification branch is responsible for predicting the object's category, while the bounding box branch focuses on determining the object's location. Each scale of detection boxes plays a distinct role in improving detection accuracy across different object sizes.

## III. METHOD DESIGN

This study improves YOLOv9s [16], as shown in Fig. 1, by incorporating the SCConv module. The SCConv module dynamically adjusts the weights of different channels to more effectively capture important spatial information while suppressing irrelevant or redundant features. The C3Ghost module is introduced, and through an optimized design, it maintains a low computational cost, allowing the model to sustain fast inference speed even when handling complex tasks. Additionally, the CARAFE upsampling operator is used to reorganize the feature map in a content-aware manner, enhancing the fine-grained reconstruction of steel surface defect regions, making defect boundaries clearer.

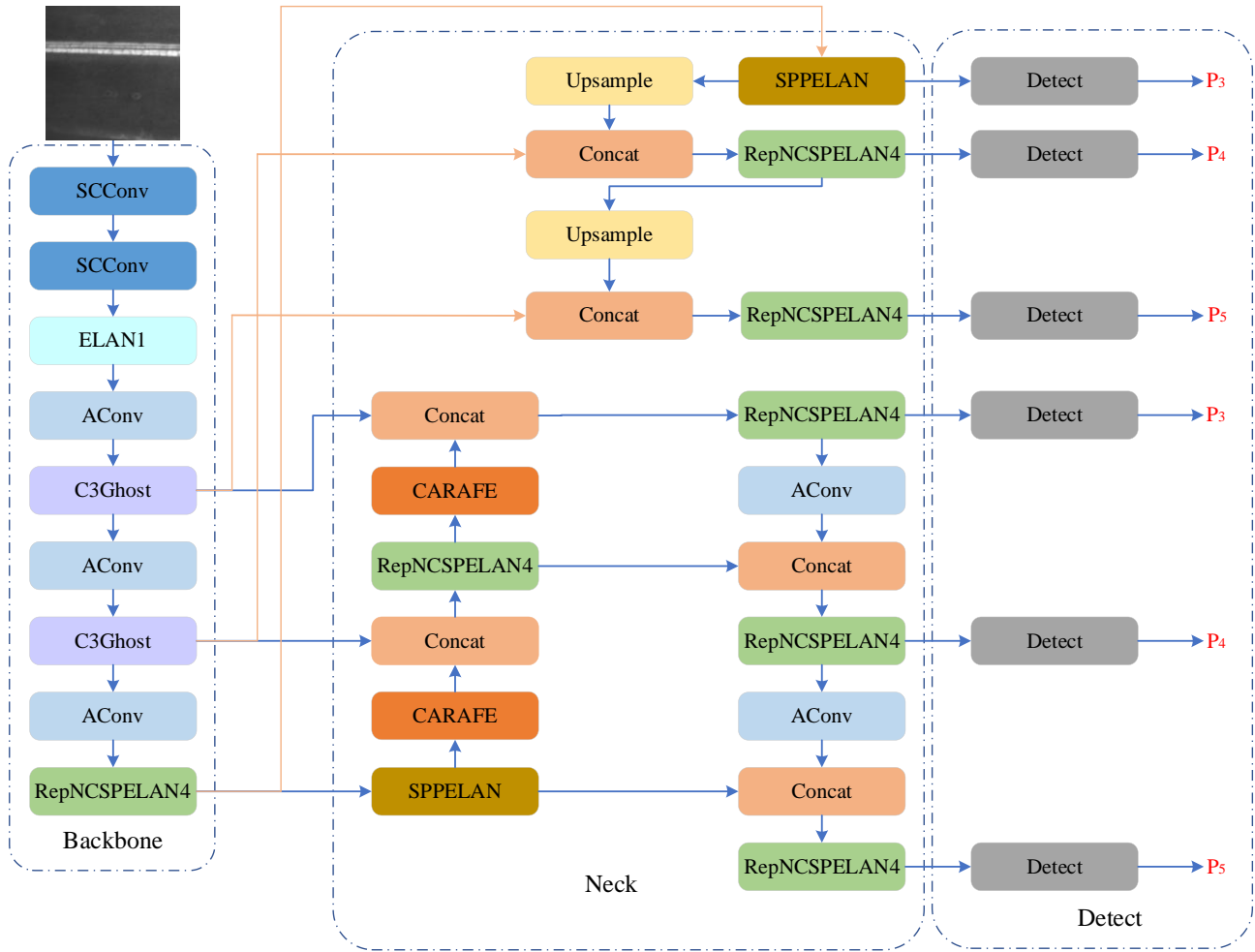

Fig. 1. Improved YOLOv9s network architecture diagram.

### A. SCConv Module

YOLOv9s, as an efficient object detection model, demonstrates outstanding performance in various complex scenarios. However, its convolutional layers still exhibit redundancy in both spatial and channel dimensions during feature extraction, which not only increases the computational burden but also limits the efficiency of feature representation to some extent. To address this issue, we introduce the SCConv [17] (Spatial and Channel Reconstruction Convolution) module, which optimizes the features through the Spatial Reconstruction Unit (SRU) and Channel Reconstruction Unit (CRU), thereby reducing redundant computations and enhancing feature representation capability.

As shown in Fig. 2, to further elaborate on the principles and process of the SRU structure, we can delve deeper into the details of the separation and reconstruction operations. The core objective of the separation operation is to distinguish between feature maps with rich information and those with less information. In this way, the SRU can extract useful information while reducing reliance on irrelevant data.

First, the input feature map $X \in R(N \times C \times H \times W)$ is processed through normalization. This is done using the Group Normalization (GN) layer. This normalization operation helps to eliminate biases between different batches of data in convolutional neural networks, making the model training more stable. The normalization formula is [Formula (1)]:

$$X_{out} = GN(X) = \frac{\gamma(X - \mu)}{\sqrt{\sigma^2 + \varepsilon}} + \beta \qquad (1)$$

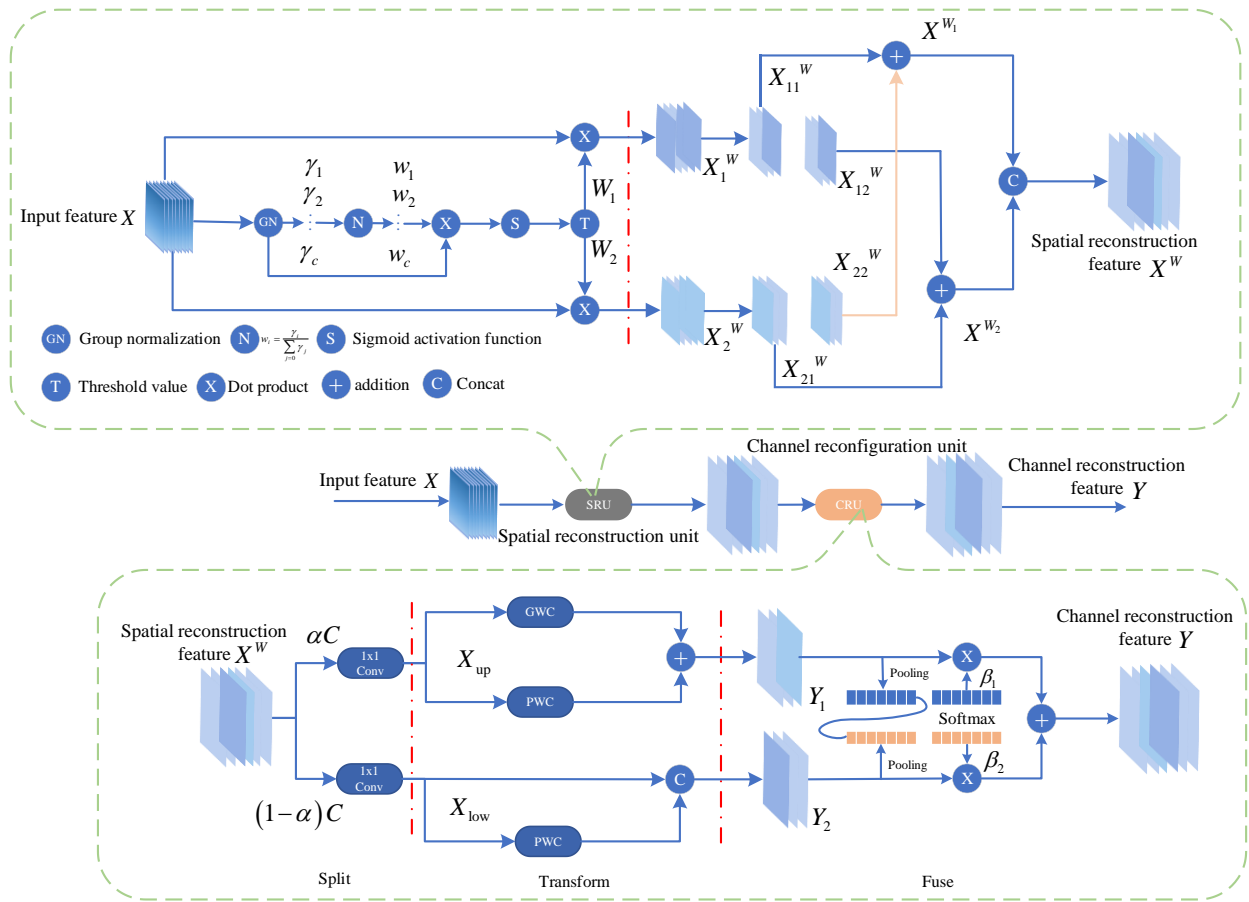

Fig. 2. SCConv module.

Here, $\mu$ and $\sigma$ are the mean and standard deviation of the input feature map $X$ along a specific dimension (usually the channel dimension). $\gamma$ and $\beta$ are the trainable affine transformation parameters, and $\varepsilon$ is a small constant used to prevent division by zero when the standard deviation is zero.

Next, we calculate the weight $W_\gamma$ based on each channel feature, which aims to measure the importance of each channel in the overall feature map. The calculation of weight $W_\gamma$ is as follows [see Formula (2)]:

$$W_\gamma = \{w_i\} = \frac{\gamma_i}{\sum_{j=1}^{C} \gamma_j}, \quad i,j = 1,2,\ldots,C \quad (2)$$

Here, $\gamma_i$ represents the affine transformation parameters of each channel in group normalization. The normalization of the weights is performed by dividing each channel's $\gamma_i$ by the sum of the affine parameters of all channels, ensuring that the weight values remain within a reasonable range.

Next, the normalized weight $W_\gamma$ is gated through the Sigmoid activation function, yielding a value within the range (0, 1), which represents the "importance" of each channel feature. Specifically as in Formula (3):

$$W = \text{Gate}(\sigma(W_\gamma(GN(X)))) \quad (3)$$

Here, $\sigma(\cdot)$ is the Sigmoid activation function, which restricts the output to the range (0, 1). Then, a threshold is set (usually 0.5), and the weights greater than this threshold are set to 1 to obtain information-rich weights $W_1$; weights smaller than the threshold are set to 0 to obtain information-poor weights $W_2$.

Subsequently, the input feature $X$ is mapped through two different weight matrices $w_1$ and $w_2$, resulting in information-rich feature $X_1^w$ and information-poor feature $X_2^w$. Then, these two types of features are fused through a cross-addition method to further enhance the expressive power of feature information and reduce spatial redundancy. Finally, the fused features $X^{w1}$ and $X^{w2}$ are concatenated together to obtain the final spatially reconstructed feature map $X^w$, which is represented by the following Formula (4):

$$\begin{cases} X_1^W = W_1 \otimes X, \\ X_2^W = W_2 \otimes X, \\ X_{11}^w \oplus X_{22}^w = X^{w1}, \\ X_{21}^w \oplus X_{12}^w = X^{w2}, \\ X^{w1} \bigcup X^{w2} = X^w. \end{cases} \quad (4)$$

In the formula, $\otimes$ represents element-wise multiplication, $\oplus$ represents element-wise addition, and $\bigcup$ represents the Concat module.

As shown in Fig. 2, the CRU (Channel Reduction Unit) structure is discussed in terms of how it optimizes feature maps through a series of operations to reduce redundancy and improve computational efficiency. The three main stages of CRU include Split, Transform, and Fuse.

In the Split stage, the feature map $X^w$ is divided along the channel dimension into two parts: the upper part contains $\alpha C$ channels, and the lower part contains $(1-\alpha)C$ channels, where $\alpha$ is a tunable parameter, typically set to 0.5, indicating that the feature map is split into two equal parts. The purpose of this step is to reduce the number of channels in each part, thereby reducing the computational burden. Next, a 1x1 convolution operation is applied to compress the number of channels in the feature map. The 1x1 convolution has the ability to reduce the number of channels while also serving as a tool for feature transformation. The compression ratio r controls the number of channels in the output feature map, balancing the trade-off between computational cost and performance. The compressed feature map is then divided into the upper part $X_{up}$ and the lower part $X_{low}$, preparing for the next stage of operations.

The Transform stage is the core part of the CRU, primarily aimed at extracting features and enhancing computational efficiency through two operations: In the upper transformation stage, the input is the upper part of the feature map $X_{up}$ obtained from the split. Groupwise Convolution (GWC) and Pointwise Convolution (PWC) are used for feature extraction. Groupwise Convolution divides the input channels into several small groups for convolution operations, thereby reducing computational complexity. Pointwise Convolution performs a $1 \times 1$ convolution at each pixel location to further extract features. The formula for the upper transformation is [Formula (5)]:

$$Y_1 = M^G X_{up} + M^{P_1} X_{up} \quad (5)$$

Here, $M^G$ and $M^{P_1}$ are the learnable weights of GWC and PWC, respectively, and $Y_1$ is the output feature map of the upper transformation stage.

In the lower transformation stage, the input is the lower part of the split feature map $X_{low}$. PWC is used to extract features, and the result is concatenated with the original $X_{low}$ feature map. The concatenation operation is denoted by the symbol $\bigcup$. The formula for the lower transformation is [Formula (6)]:

$$Y_2 = M^{P_2} X_{low} \bigcup X_{low} \quad (6)$$

Here, $M^{P_2}$ is the learnable weight of PWC, and $Y_2$ is the output of the lower transformation stage.

In the Fusion stage, the simplified SKNet method is used to merge the output features $Y_1$ and $Y_2$ from the upper and lower transformation stages. Global spatial information is obtained through global average pooling. The formula for the global channel descriptor $S_m$ is as follows [Formula (7)]:

$$S_m = \frac{1}{H \times W} \sum_{i=1}^{H} \sum_{j=1}^{W} Y_c(i,j), m=1,2 \quad (7)$$

Subsequently, the two global channel descriptors $S_1$ and $S_2$ are stacked together.

Channel attention operations are used to generate importance weights $\beta_1$ and $\beta_2$, which represent the importance of the corresponding channels. The calculation formula is as follows [Formula (8)]:

$$\beta_1 = \frac{e^{s_1}}{e^{s_1}+e^{s_2}}, \beta_2 = \frac{e^{s_2}}{e^{s_1}+e^{s_2}} \quad (8)$$

Finally, the reconstructed feature $\beta_1 Y_1 + \beta_2 Y_2$ is obtained through the weighted calculation.

*B. C3Ghost Module*

In the YOLOv9s model, the main goal of replacing the RepNCSPELAN4 module with the C3Ghost [15,18] module is to enhance the model's computational efficiency and detection performance. The core idea behind the design of the C3Ghost module is to optimize the network structure, reduce computational load and the number of parameters, while not sacrificing the model's detection accuracy, thereby achieving efficient object detection. Compared to traditional convolutional modules, the C3Ghost module uses the GhostConv operation, which can extract more effective features at a lower computational cost. Therefore, we introduced the C3Ghost module to significantly improve computational efficiency while maintaining detection accuracy.

As shown in Fig. 3, the workflow of the C3Ghost module is as follows: First, the input feature map X is processed through a convolution layer (Conv) for preliminary processing. Then, it passes through multiple GhostConvBottleneck modules, which improve efficiency by reducing computational load and the number of parameters, while enhancing feature extraction capabilities. After that, a Concat operation is performed, followed by a Conv operation from another path. Finally, the concatenated feature map undergoes a convolution layer (Conv) to generate the final output X.

*C. CARAFE Upsampling Operator*

The CARAFE [19] upsampling operator is introduced in the YOLOv9s model to replace traditional upsampling methods, primarily due to its unique advantages. CARAFE is a content-aware upsampling mechanism that dynamically adjusts the upsampling process based on the input features, generating higher-quality feature maps, especially excelling in small object detection. Therefore, using the CARAFE upsampling operator can improve the model's detection accuracy for small objects and overall performance.

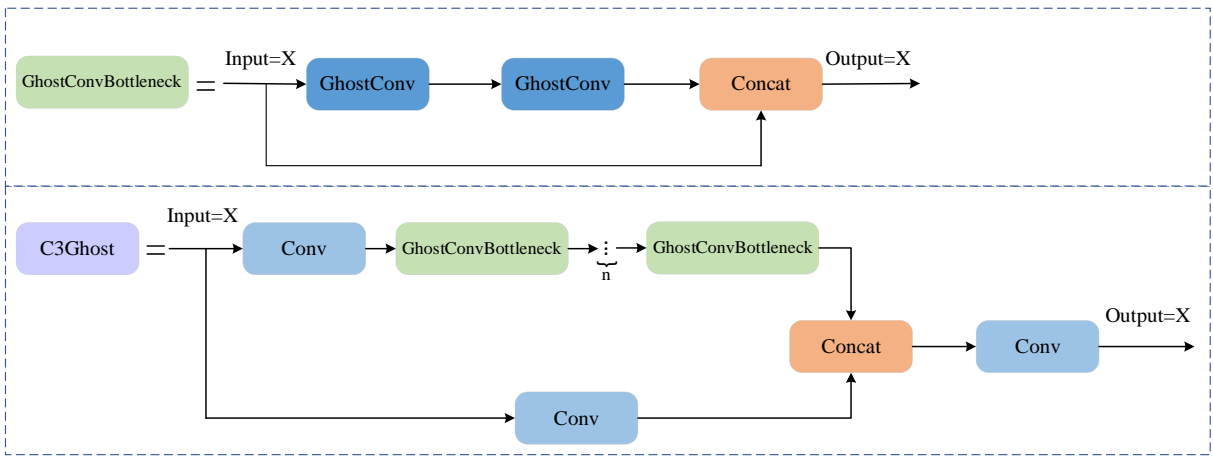

Fig. 3. C3Ghost module.

As shown in Fig. 4, in the upsampling kernel prediction section, first, the channel compression part compresses the input feature map X using a weight matrix $W$, reducing the number of channels from $C$ to $C_m$, resulting in a compressed feature representation. Next, the feature map undergoes content encoding, converting the input feature map into a more processable form. Then, a spatial dimension expansion operation is performed to obtain an upsampling kernel of the form $\sigma H \times \sigma W \times k_{up}^2$, where the upsampling factor $\sigma$ is set. Finally, a kernel normalization operation is applied to ensure that the sum of the kernel weights equals 1, resulting in the output feature map O.

In the feature reorganization section, the feature map undergoes spatial dimension expansion, converting into feature blocks of $K_{up} \times K_{up}$. These feature blocks will prepare for the final upsampling operation. Finally, a pointwise operation is performed, where the expanded feature blocks are corresponded with the points in the original feature map, generating the final upsampled feature map $X'$.

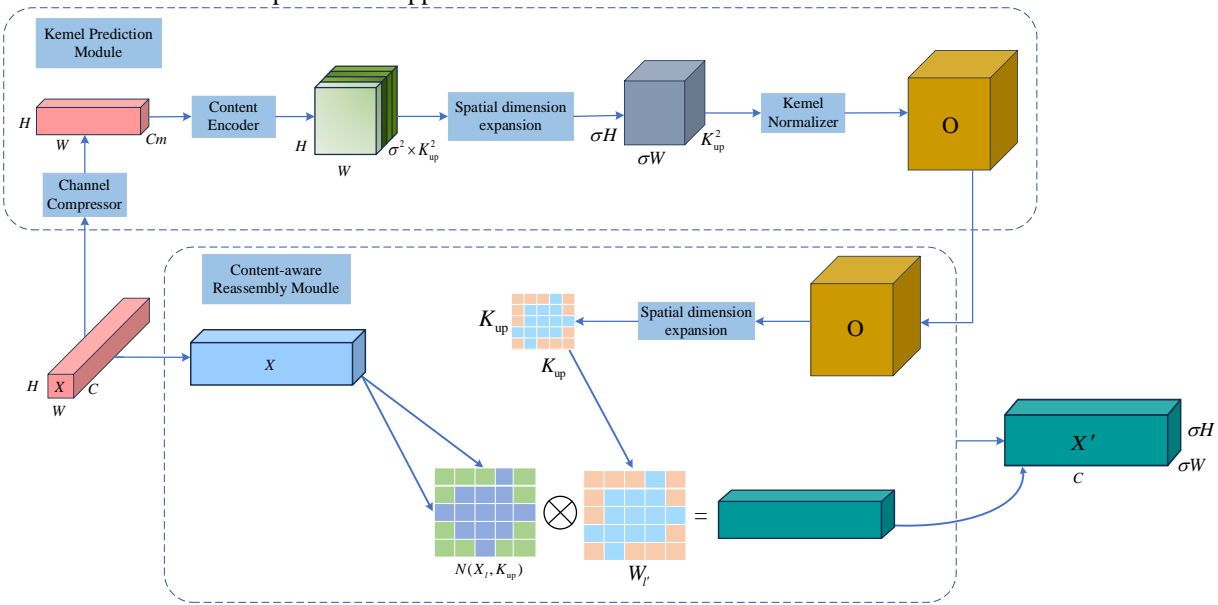

Fig. 4. CARAFE Upsampling operator.

## IV. EXPERIMENTAL SETUP

### A. Dataset

The steel surface defect dataset used in this study, provided by Northeastern University, contains six typical defect categories: a) Crazing, b) Inclusion, c) Patches, d) Pitted Surface, e) Rolled-in Scale, and f) Scratches. The dataset consists of 1,800 images, with 300 samples for each category, ensuring a balanced distribution that comprehensively covers common defect types and features found on steel surfaces. These images were captured using industrial equipment and authentically reproduce the visual characteristics and complexity of surface defects in real-world production processes. Each image is annotated with bounding boxes, a common format used in object detection tasks, which include the defect location and corresponding category information. To reasonably allocate the dataset for training and testing purposes, it is divided into training, validation, and test sets in a 7:1:2 ratio, containing 1,260, 180, and 360 images, respectively. This division ensures that the training phase has sufficient data to effectively learn the features of each defect category, while the validation and test sets provide a scientific evaluation of the model's performance. The six types of steel surface defect images are shown in Fig. 5.

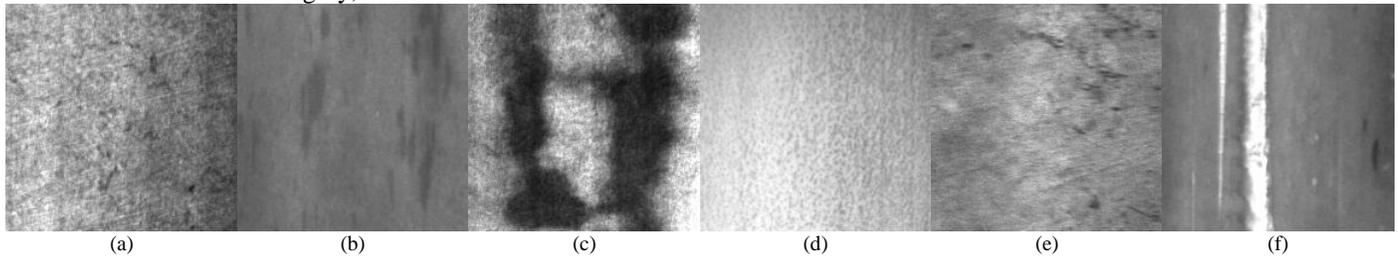

(a)　　　　　(b)　　　　　(c)　　　　　(d)　　　　　(e)　　　　　(f)

Fig. 5. Six types of steel surface defects.

### B. Experimental Platform

The experiments in this study were conducted on a high-performance computing platform with comprehensive software and hardware configurations, which fully meet the computational resource requirements of deep learning tasks. The software environment uses the Ubuntu 20.04 operating system, with model development and training based on Python 3.8 and PyTorch 1.11.0, and it supports CUDA 11.3 acceleration, significantly improving the efficiency of large-scale data processing during training. The hardware environment includes an NVIDIA RTX 4090 GPU (24GB memory), providing powerful support for deep learning computations; a 22-core AMD EPYC 7T83 64-Core Processor (vCPU), suitable for data preprocessing and training task scheduling; 90GB of RAM, ensuring stable loading of high-resolution images and large-scale model training; as well as a 30GB system disk and a 50GB data disk, used for operating system operation and experimental data storage, respectively. This platform adopts GPU and CPU parallel acceleration to fully utilize computational resources, improving experimental efficiency and task processing capabilities. Additionally, its software and hardware combination offer high stability and scalability, supporting the optimization and expansion needs of future experiments.

### C. Hyperparameter Settings

This study made reasonable settings for the training parameters to maximize model performance while ensuring efficient training. During the training process, the pre-trained model "yolov9-s.pt" was loaded, and the image resolution was set to 640×640. The batch size was set to 8 to adapt to hardware resources and ensure the stability of gradient updates. The number of epochs was set to 300 to ensure that the model fully learns the feature patterns in the data. The number of workers was set to 4 to parallelize the data loading process, thereby improving training efficiency. In terms of optimizer parameters, the learning rate was set to 0.01, which provides a good balance between model convergence speed and training stability. The momentum parameter was set to 0.937 to reduce gradient oscillation and accelerate the convergence process. The weight decay coefficient was set to 0.0005 to limit the model's complexity and prevent overfitting. With these fine-tuned parameter settings, this study ensured both the efficiency of the training process and the stability of model performance.

### D. Evaluation Metrics

The experiments in this study use F1 score, Precision (P), Recall (R), Average Precision (AP), and mean Average Precision (mAP) as evaluation metrics [20], and also consider the number of parameters (Parameters). The calculation formulas for these metrics are as follows [Formula (9) to (13)]:

$$\text{Precision} = \frac{T_P}{T_P + F_P} \quad (9)$$

$$\text{Recall} = \frac{T_P}{T_P + F_N} \quad (10)$$

$$\text{AP} = \int_0^1 P(R)\,dR \quad (11)$$

$$\text{mAP} = \frac{1}{n}\sum_{i=0}^{n} AP(i) \quad (12)$$

$$F1 = \frac{2 \times \text{Precision} \times \text{Recall}}{\text{Precision} + \text{Recall}} \quad (13)$$

Here, $T_P$ represents the number of correctly detected defect targets; $F_p$ represents the number of incorrectly detected defect targets; $F_N$ represents the number of missed defect targets; n represents the number of defect categories; and $AP(i)$ represents the average precision for the i-th target class.

## V. EXPERIMENTAL ANALYSIS

### A. Algorithm Comparison Experiment

To validate the proposed performance improvements, we compared the improved YOLOv9s model (Ours) to RT-DETR-18, Faster RCNN, SSD, and seven other mainstream YOLO target detection models, all tested on the same data set and training rounds. The results are shown in Table I. The comparison metrics include Precision, Recall, mean Average Precision (mAP@0.5), Frames Per Second (FPS), computational complexity (GFLOPS), and model parameter count (Params). These metrics enable us to evaluate the performance of each algorithm and its efficiency under different computational resources. Among all the algorithms, the proposed improved model demonstrates a superior balance between precision and recall, while also achieving significant improvements in mAP and inference speed.

TABLE I. EXPERIMENTAL COMPARISON RESULTS OF DIFFERENT ALGORITHMS

| Algorithm | Precision/% | Recall/% | mAP@0.5/% | FPS (f/s) | GFLOPS | Params (M) |
|---|---|---|---|---|---|---|
| Faster R-CNN[21] | 79.9 | —— | 77.5 | —— | —— | —— |
| SSD[22] | 70.2 | —— | 68.6 | —— | —— | —— |
| YOLOv5s[23] | 71.7 | 71.2 | 75.2 | 208.3 | 15.8 | 7.02 |
| YOLOv7[24] | 63.3 | 74.8 | 71.7 | 109.8 | 103.2 | 36.50 |
| YOLOv8s[25] | 75.7 | 67.1 | 71.7 | 312.5 | 28.4 | 11.12 |
| YOLOv9s | 70.7 | 77.3 | 78.1 | 94.3 | 38.7 | 9.60 |
| YOLOv10s[26] | 77.5 | 67.4 | 74.4 | 125.0 | 24.5 | 8.03 |
| YOLOv11s[27] | 74.1 | 74.7 | 78.4 | 86.2 | 21.3 | 9.41 |
| RT-DETR-18[28] | 79.3 | 67.4 | 73.4 | 68.0 | 57.0 | 19.87 |
| Ours | **77.0** | **72.4** | **79.6** | **109.9** | **36.9** | **9.07** |

Table I presents the comparison experiment results of various object detection algorithms, including Precision, Recall, mean Average Precision (mAP@0.5), Frames Per Second (FPS), computational complexity (GFLOPS), and model parameter count (Params). Among these algorithms, The Precision of Faster R-CNN reached 79.9%, and the mAP@0.5 achieved 77.5%. The Precision of SSD reached 70.2%, and the mAP@0.5 achieved 68.6%. YOLOv5s has a high inference speed (208.3 FPS) and low computational complexity (15.8 GFLOPS), but its Precision (71.7%) and Recall (71.2%) are relatively moderate. YOLOv7 performs well in Recall (74.8%), but its Precision is lower (63.3%) and computational complexity is higher. YOLOv8s achieves a high Precision (75.7%) but has a lower Recall (67.1%) and an extremely fast inference speed (312.5 FPS). YOLOv9s performs well in Recall and mAP (78.1%), but its inference speed is slower (94.3 FPS). YOLOv10s has high Precision (77.5%) with moderate inference speed (125.0 FPS). YOLOv11s shows a high Recall (74.7%) but slower inference speed (86.2 FPS). RT-DETR-18, although having high Precision (79.3%), has slower inference speed and lower mAP (73.4%). Our model achieves a good balance in both Precision (77.0%) and Recall (72.4%), with an mAP of 79.6%, inference speed of 109.9 FPS, moderate computational complexity (36.9 GFLOPS), and a model parameter count of 9.07 M. It demonstrates high overall performance, making it an efficient and balanced choice.

TABLE II. COMPARISON OF VARIOUS ALGORITHMS IN TERMS OF AVERAGE PRECISION (AP%)

| Category \ Algorithm | Inclusion | Scratches | Crazing | Patches | Rolled-in_Scale | Pitted_Surface |
|---|---|---|---|---|---|---|
| YOLOv5s | 78.6 | 91.1 | 41.3 | 89.1 | 70.9 | 80.3 |
| YOLOv7 | 75.6 | 80.5 | 38.9 | 89.9 | 66.6 | 78.9 |
| YOLOv8s | 70.2 | 89.0 | 32.8 | 86.4 | 66.1 | 85.7 |
| YOLOv9s | 81.2 | 94.3 | 48.0 | 89.1 | 70.8 | 85.5 |
| YOLOv10s | 75.8 | 90.5 | 43.8 | 81.7 | 74.0 | 80.7 |
| YOLOv11s | 82.1 | 92.3 | 49.0 | 89.9 | 69.5 | 87.8 |
| RT-DETR-18 | 78.6 | 93.1 | 48.9 | 88.0 | 52.5 | 79.5 |
| Ours | **84.1** | **92.5** | **52.0** | **88.8** | **75.0** | **85.1** |

Based on the Average Precision (AP) comparison results in Table II, the "Ours" model performs the best across multiple defect categories, particularly excelling in the inclusion and crazing categories, with precision rates of 84.1% and 52.0%, respectively, significantly higher than all other algorithms. YOLOv9s and YOLOv11s also perform well in most categories, especially in scratches, where they achieve precision rates of 94.3% and 92.3%, respectively, although they are slightly lacking in the Rolled-in_scale category. YOLOv5s, YOLOv7, and YOLOv10s perform well in the scratches and patches categories, but their precision in crazing and Rolled-in_scale is lower. YOLOv8s shows strong performance in scratches and Pitted_surface, but lags significantly in crazing. RT-DETR-18 also demonstrates good performance in scratches and crazing. In summary, the "Ours" model outperforms other algorithms across multiple defect categories, showing significant performance improvements.

*B. Result Visualization*

To evaluate the effectiveness of the improved algorithm, it needs to be compared with the original algorithm. YOLOv9s, as a classic object detection algorithm, is widely used in various tasks. With continuous optimization of the algorithm, the improved YOLOv9s has shown better performance in both precision and recall. The F1 score, which considers both precision and recall, provides an effective way to assess the algorithm's performance in detection tasks. By comparing the F1 scores of YOLOv9s and the improved YOLOv9s, we can visually understand the improvements in the balance between precision and recall, thereby evaluating the performance optimization effect of the improved algorithm.

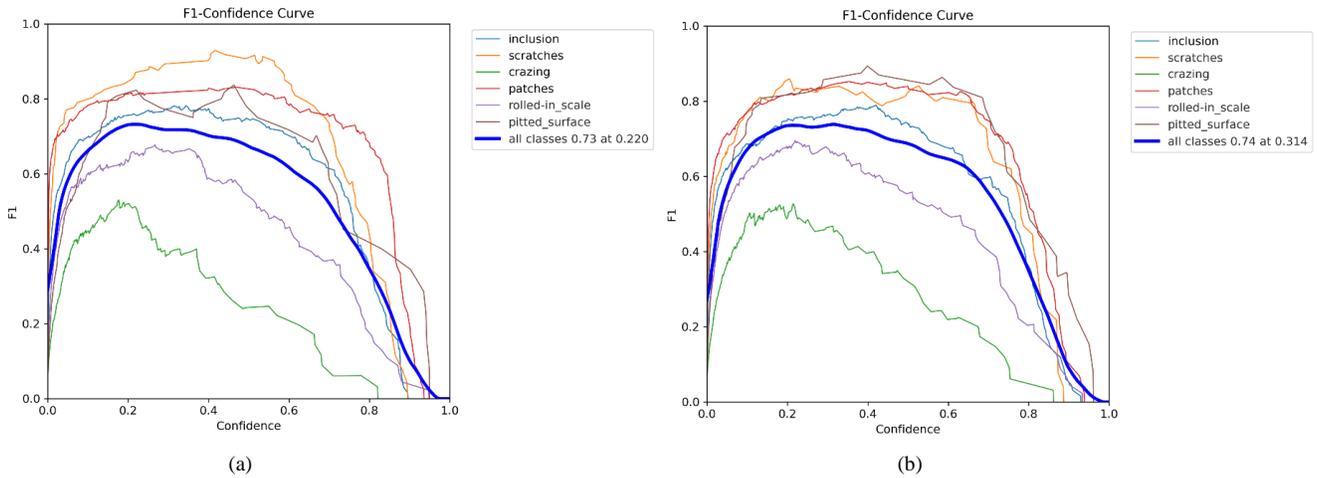

Fig. 6. F1-Confidence Curve: (a) YOLOv9s Algorithm; (b) Improved YOLOv9s Algorithm.

Fig. 6 displays the F1-Confidence curves for both YOLOv9s and the improved YOLOv9s. The blue curve represents the F1 score for all categories. From the comparison, it is evident that the improved YOLOv9s shows a significant enhancement in F1 scores across categories, particularly in the higher confidence regions, indicating a clear optimization in precision. Specifically, in the left-side YOLOv9s plot, the blue curve reaches an F1 score of 0.73 at a confidence of 0.220, while in the right-side improved YOLOv9s plot, the blue curve reaches an F1 score of 0.74 at a confidence of 0.314. This demonstrates that the improved YOLOv9s has an overall increase in F1 score across all categories, with better performance at higher confidence levels.

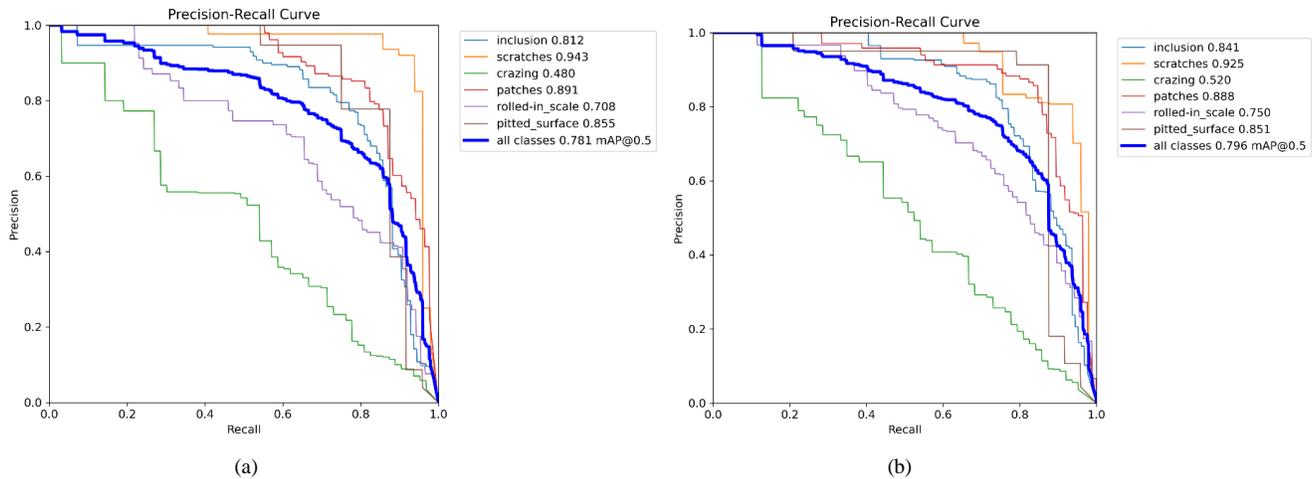

Fig. 7. Precision-Recall Curve: (a) YOLOv9s Algorithm; (b) Improved YOLOv9s Algorithm.

Fig. 7 shows the Precision-Recall (PR) curves for YOLOv9s and the improved YOLOv9s across different categories. In the left-side YOLOv9s plot, the blue curve represents the Precision-Recall curve for all categories, with an mAP value of 0.781. In the right-side plot of the improved YOLOv9s, the blue curve has an mAP value of 0.796, which shows an improvement over YOLOv9s, indicating that the improved algorithm performs better in balancing precision and recall. Overall, the improved YOLOv9s shows enhanced precision across multiple categories, particularly in the "inclusion" and "crazing" categories, where the improved algorithm demonstrates superior performance.

Fig. 8 shows the Precision-Confidence curves for YOLOv9s and the improved YOLOv9s across different categories. In the left-side YOLOv9s plot, the blue curve represents the Precision-Confidence curve for all categories, with a precision of 1.00 at a confidence of 0.901. In the right-side plot of the improved YOLOv9s, the blue curve also reaches a precision of 1.00, but at a confidence of 0.921. A comparison reveals that the improved YOLOv9s shows an overall increase in precision at higher confidence levels and outperforms YOLOv9s in multiple categories, especially demonstrating more stable precision at lower confidence levels.

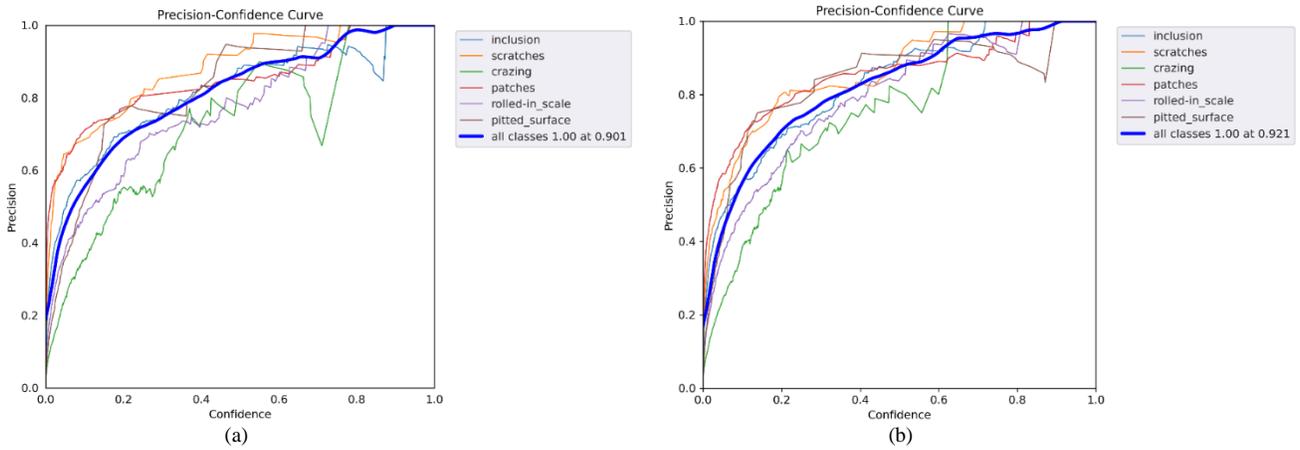

Fig. 8. Precision-Confidence Curve: (a) YOLOv9s Algorithm; (b) Improved YOLOv9s Algorithm.

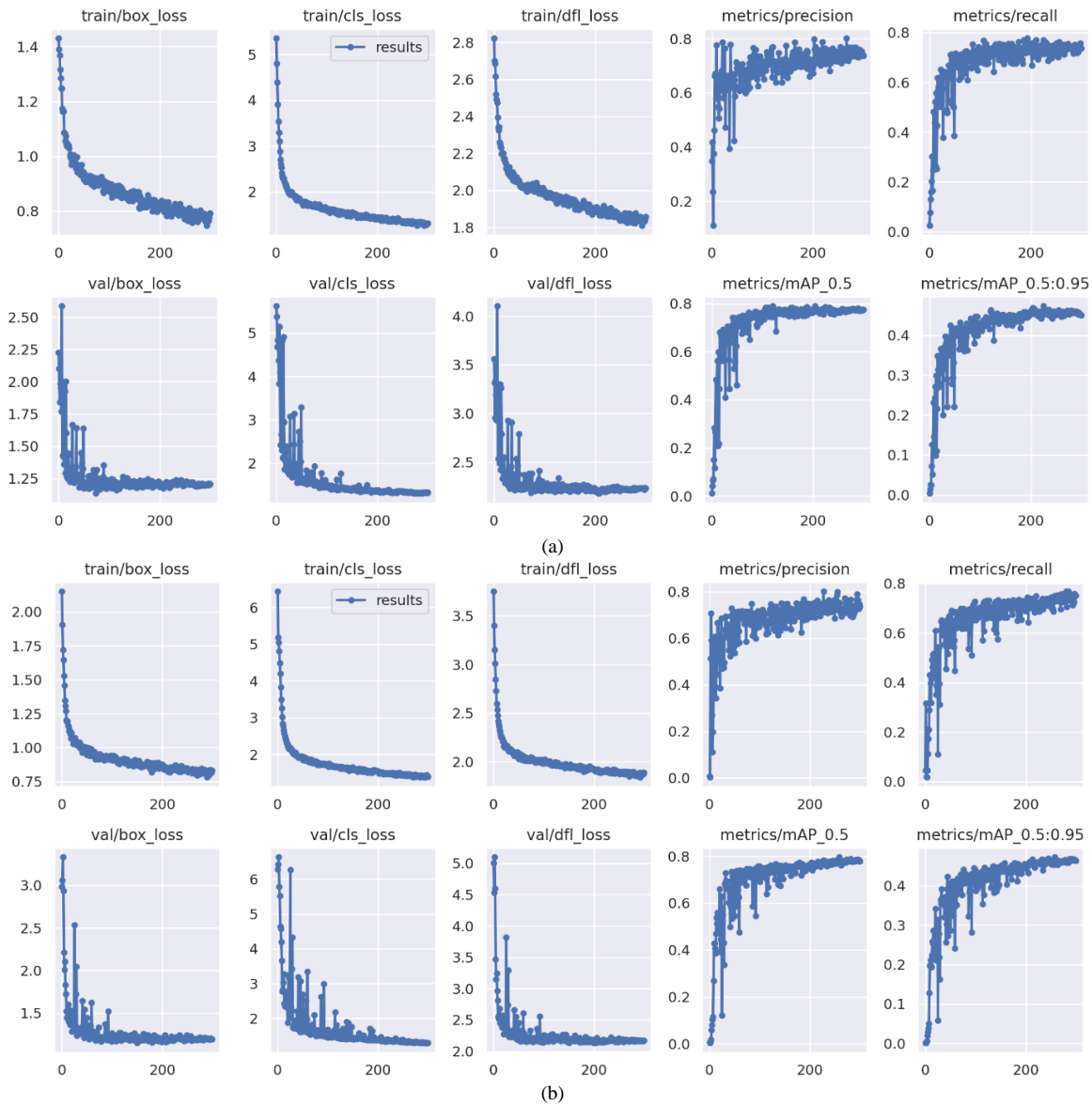

Fig. 9. Change Curves of Various Loss Functions and Evaluation Metrics During Training and Validation: (a) YOLOv9s Algorithm; (b) Improved YOLOv9s Algorithm.

Fig. 9 shows the change curves of various loss functions and evaluation metrics during the training and validation process for YOLOv9s and the improved YOLOv9s. These include box loss, classification loss, depth regression loss, precision, recall, and mean average precision (mAP). It can be observed that as training progresses, the losses for both the training and validation sets gradually decrease, indicating continuous improvement in the model's performance in box prediction, classification, and depth regression. Meanwhile, the precision, recall, and mAP metrics also increase, showing a significant improvement in the detection capabilities of the improved YOLOv9s. Overall, the improved YOLOv9s outperforms YOLOv9s across all metrics, demonstrating effective optimization and stronger detection performance during training.

As shown in Fig. 10 below, we compare the detection results of various algorithms on the steel surface defect dataset, including a) Faster R-CNN, b) SSD, c) YOLOv5s, d) YOLOv7, e) YOLOv8s, f) YOLOv9s, g) YOLOv10s, h) YOLOv11s, i) RT-DETR-18, and j) Ours. Through intuitive visualization, we can clearly observe the recognition performance of each algorithm on different targets, including localization accuracy and classification confidence. The improved algorithm (Ours) demonstrates higher localization accuracy across multiple defect categories, showing a distinct advantage over the other algorithms. These comparisons allow us to more comprehensively assess the performance improvements of the improved algorithm in practical applications.

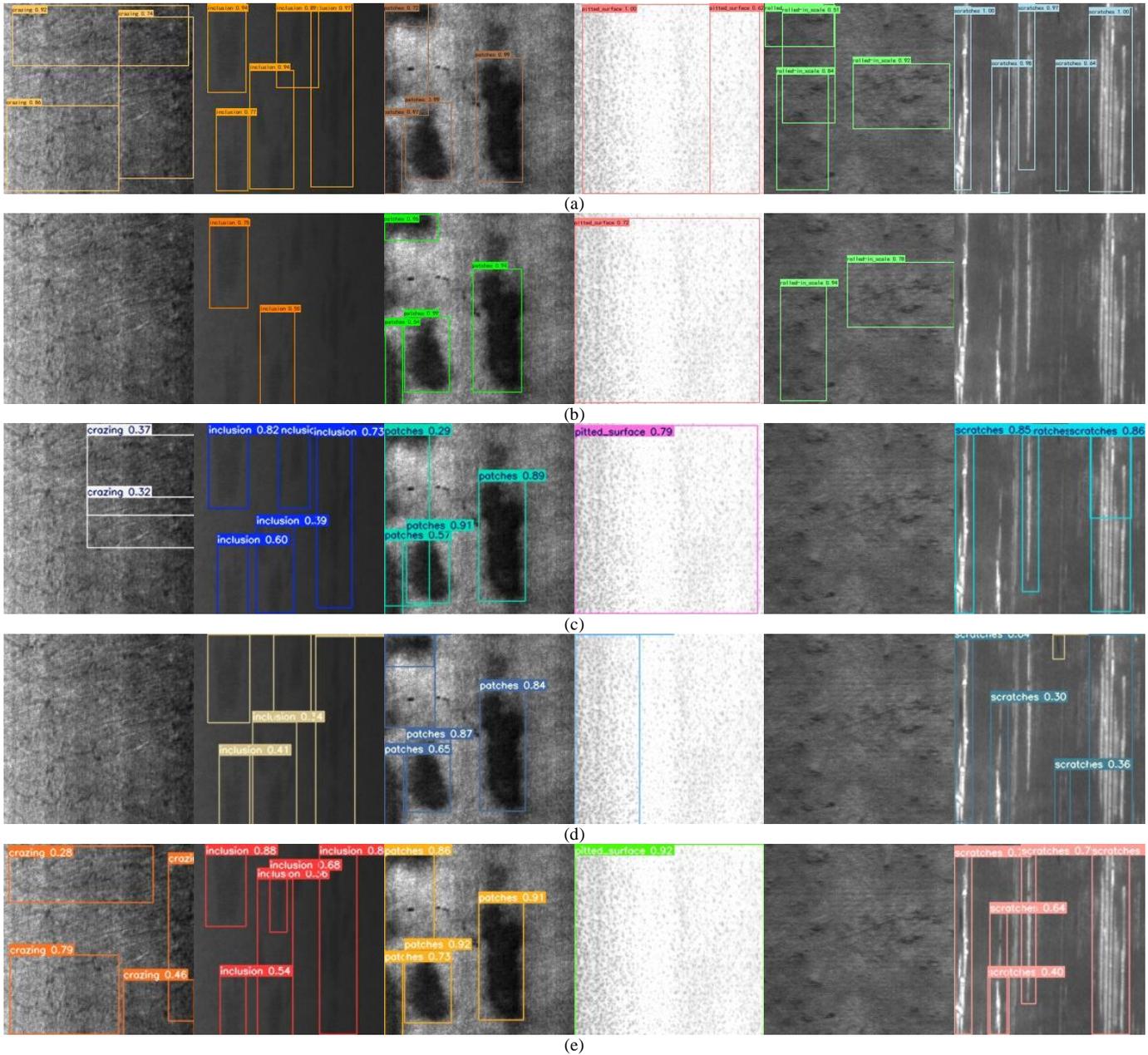

(a)

(b)

(c)

(d)

(e)

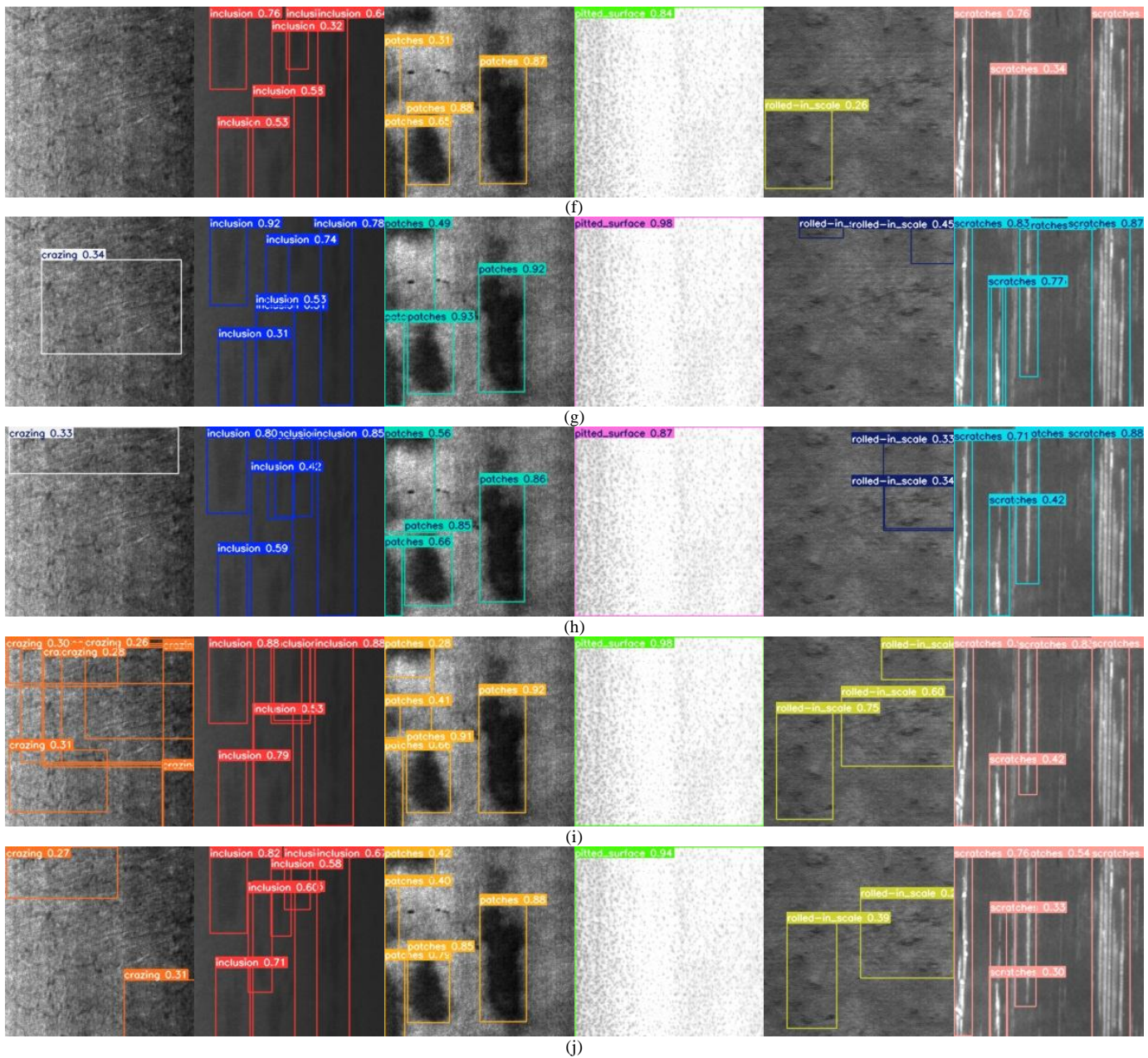

Fig. 10. Detection Results on the Steel Surface Defect Dataset: (a) Faster RCNN; (b) SSD; (c) YOLOv5s; (d) YOLOv7; (e) YOLOv8s; (f) YOLOv9s; (g) YOLOv10s; (h) YOLOv11s; (i) RT-DETR-18; (j) Ours.

Based on the visualization results of steel surface defect detection shown in Fig. 10, different models exhibit varying advantages and disadvantages when detecting defects such as crazing, inclusion, patches, pitted surface, rolled-in scale, and scratches. Overall, traditional object detection methods like Faster R-CNN and SSD have certain limitations. Although Faster R-CNN produces more detection boxes, it suffers from severe target overlapping, making small targets prone to being missed. SSD performs well in detecting large targets but has weaker recognition capabilities for small targets, resulting in a relatively low overall recall rate. In comparison, YOLOv5s demonstrates a noticeable improvement in both recall rate and precision over SSD, although some instances of missed detection still occur. YOLOv7 performs slightly worse than YOLOv5s, with a higher false detection rate and an unremarkable recall rate. YOLOv8s performs well in detecting large and medium-sized targets but still struggles with certain small targets. YOLOv9s and YOLOv10s further enhance detection confidence and recall rates, showing improved detection performance for rolled-in scale and scratches while reducing false detections. YOLOv11s achieves the best overall performance among all categories, offering higher recall rates and confidence scores, more precise detection boxes, and the lowest false detection rate. Additionally, RT-DETR-18 achieves a well-balanced trade-off between precision and recall. Compared to the YOLO series, it provides higher detection confidence and a lower false detection rate, particularly demonstrating stable performance in detecting pitted surfaces and patches. However, there is still room for improvement in detecting certain small targets, such as scratches. Ultimately, the proposed model (Ours) outperforms all other detection models, delivering the most precise detection boxes, the highest recall rate, the most stable confidence scores, and the lowest false detection rate. Compared with other models, it accurately identifies all defect categories, effectively avoiding common issues of missed and false detections. Notably, it shows significant advantages in detecting small targets such as scratches and inclusions, highlighting its high applicability and reliability in steel surface defect detection tasks. Future optimizations could focus on enhancing detection capabilities for extremely small targets (e.g., fine scratches) to improve detection sensitivity, making it even more efficient and precise for industrial inspection applications.

### C. Ablation Study

To further validate the impact of different modules on model performance, we conducted ablation experiments. The ablation study compares the performance of different model combinations across multiple performance metrics, including Precision, Recall, mean Average Precision (mAP), Frames Per Second (FPS), GFLOPS, and parameter count (Params). These experiments aim to assess the contribution of each module to the model's performance, helping us understand which optimizations effectively improve the detection results.

TABLE III. ABLATION STUDY RESULTS

| Number | Experiment | Precision/% | Recall/% | mAP@0.5/% | FPS (f/s) | GFLOPS | Params (M) |
|---|---|---|---|---|---|---|---|
| 1 | YOLOv9s | 70.7 | 77.3 | 78.1 | 94.3 | 38.7 | 9.60 |
| 2 | YOLOv9s+CARAFE | 76.4 | 71.3 | 78.3 | 82.6 | 39.0 | 9.74 |
| 3 | YOLOv9s+C3Ghost | 78.4 | 70.3 | 79.5 | 131.5 | 34.5 | 8.91 |
| 4 | YOLOv9s+SCConv | 73.9 | 75.5 | 79.1 | 75.7 | 40.8 | 9.62 |
| 5 | YOLOv9s+CARAFE+C3Ghost+SCConv | **77.0** | **72.4** | **79.6** | **109.9** | **36.9** | **9.07** |

According to the ablation experiment results in Table III, each model shows different performance across various metrics such as precision, recall, mAP, FPS, GFLOPS, and parameter count. Experiment 1 uses YOLOv9s as the baseline model, with a precision of 70.7%, recall of 77.3%, mAP of 78.1%, and FPS of 94.3, demonstrating good performance. Experiment 2 introduces the CARAFE module based on Experiment 1, increasing precision to 76.4%, but recall drops to 71.3%, and FPS significantly decreases to 82.6. Experiment 3 adds the C3Ghost module to Experiment 1, resulting in a further improvement in precision to 78.4% and a slight decrease in recall to 70.3%. Additionally, the FPS notably increases to 131.5, reflecting significant speed optimization. Experiment 4 introduces the SCConv module based on Experiment 1, improving both precision and recall, but FPS drops to 75.7. Finally, Experiment 5 combines the CARAFE, C3Ghost, and SCConv modules, showing improvements in precision, recall, and mAP, with precision at 77.0%, recall at 72.4%, mAP at 79.6%, and FPS rising to 109.9, despite a slight increase in parameter count. Overall, Experiment 5 (YOLOv9s + CARAFE + C3Ghost + SCConv) achieves a better balance between precision and speed, showing significant improvements over the baseline model.

### VI. DISCUSSION

Although the improved YOLOv9s model achieves significant enhancements in defect detection performance, several limitations were observed during experimental testing. First, the model's detection accuracy slightly declines when dealing with extremely small-sized defects, such as fine scratches or micro-inclusions, particularly under conditions of strong illumination interference or highly complex surface textures. Second, while the introduction of the SCConv, C3Ghost, and CARAFE modules enhances feature representation and reconstruction capabilities, the complexity of the combined modules leads to a marginal increase in inference latency compared to the original YOLOv9s. Although the overall computational overhead remains acceptable, this factor could pose potential challenges for deployment in ultra-real-time industrial detection systems where strict latency requirements exist. In future work, optimizing the lightweight structure of the improved model to further reduce inference time without compromising detection accuracy will be explored. Additionally, strategies such as multi-scale feature enhancement, adaptive noise suppression, and transfer learning under domain adaptation settings will be investigated to improve the model's performance in detecting extremely small defects and generalizing across varied production environments.

### VII. CONCLUSION

This study aims to improve the YOLOv9s model to enhance its performance in steel surface defect detection tasks. To address the shortcomings of the existing YOLOv9s model in detecting small-sized defects on complex steel surface scenes, this study proposes a deep learning-based YOLOv9s method for steel surface defect detection by introducing the SCConv module for optimization, the design optimization of the C3Ghost module, and the CARAFE upsampling operator. Detailed experimental validation of the proposed method is provided. The main conclusions of this study are as follows: First, the adoption of the SCConv module significantly improved the feature extraction efficiency and automatically suppressed redundant and unimportant features. The introduction of the C3Ghost module, by streamlining the computation process and reducing redundant convolution calculations, effectively reduced the model's computational complexity while maintaining a high feature representation capability. To address the issues of small defects and blurred boundaries in steel surface defect images, the CARAFE (Content-Aware Receptive Field) upsampling operator was incorporated. CARAFE finely reconstructs feature maps in a content-aware manner, improving the detail reconstruction accuracy of defect regions, and showing significant advantages in the clarity of defect boundaries and the localization accuracy of small defects. Experimental results show that the improved model achieved mean average precision (mAP) of 79.6% and precision of 77.0%, which represent improvements of 1.5% and 6.3%, respectively, compared to the baseline model. The findings of this study are of significant practical value for improving the quality and production efficiency of industrial products.